# STRUCTURAL MODELING OF CROSSED ROLLER WIRE RACE BEARINGS: ANALYTICAL SUBMODEL FOR THE ROLLER-WIRE-RING SET

Iñigo Martín[1], Iker Heras[1], Ibai Coria[1], Mikel Abasolo[1], Josu Aguirrebeitia[1]

[1] Department of Mechanical Engineering, University of the Basque Country (UPV/EHU), Bilbao, Spain.

## Abstract

Since wire bearings were patented in 1936, they have been used in applications where weight saving is a key aspect. Nevertheless, little work can be found in literature regarding their structural behaviour. In order to predict how a wire bearing reacts under load, the most feasible way is to model and analyse it via Finite Elements (FE). However, the great amount of elements needed to properly model the complex contact configuration among rollers, wires and raceways, make this approach unaffordable. In this sense, this manuscript develops an analytical submodel that represents the structural behaviour of the roller-wire-ring set, in order to include it later in a global FE model to account for ring flexibility and boundary conditions. This way, the computationally intensive FE modelling of these complex zones can be substituted by the analytical model developed in this work, and a much manageable global FE model can be used to predict the structural response of the bearing. This analytical model has demonstrated a good agreement with local FE models of the roller-wire-ring set.

## Keywords

## 1. Introduction

Wire bearings are special bearings where the rolling elements run on races machined in steel wires embedded within the rings. Fig. 1 illustrates a crossed roller wire bearing. This design allows building the rolling elements and the wires with hardened steel and the rings with lighter materials such as aluminium, carbon fibre or even plastics. According to manufacturers, weight savings up to 65% can be achieved if the rings are made of aluminium instead of steel as in conventional bearing [1]. Furthermore, their performance under shock loads and travelling vibrations is better, thus reducing raceway striation and brinelling problems [2]. Besides, when a raceway is damaged, not the whole bearing but only the damaged wire can be replaced. Due to the crossed layout of the rollers shown in Fig. 1, crossed roller bearings can carry axial loads in both directions, radial loads and tilting moments.

Wire bearings are increasingly used in defence applications (where shock loads are significant), as well as in robotics (where inertia reductions allows for higher accelerations and decelerations) and medical devices such as scanners (where travelling vibrations are a common issue).

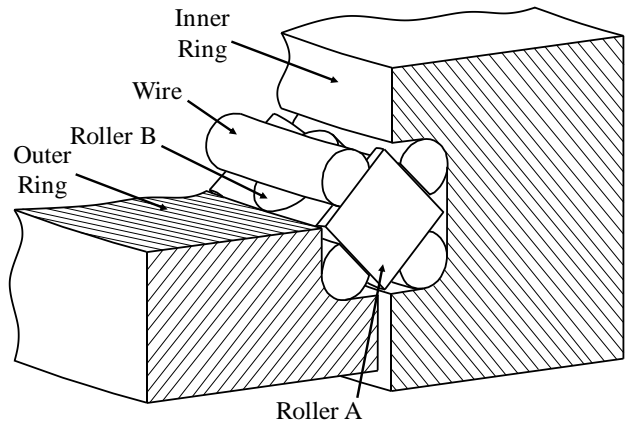

**Fig. 1.** Crossed roller wire bearing cross section.

Even though wire bearings were patented by Erich Franke in 1936, no specific standards have been developed yet and only a few works on their structural behaviour have been published. Shan et al. developed an analytical method to determine the proper preload in wire ball bearings with non-conformal ball-wire contact [3]. Gunia and Smolnicki studied the relationship between ball-wire contact stress distribution and some geometrical parameters [4]. Later, they proposed a corrected wire-raceway shape for ball bearings to reduce the stress magnitude in the most loaded areas [5]. Martin et al. compared the behaviour of conventional and wire-race ball bearings under axial loading conditions, explaining the wire-twisting phenomenon and evaluating the influence of some design parameters [6], and also developed an analytical model for the estimation of stiffness and contact results under axial load for wire-race ball bearing [7]. Aguirrebeitia et al. obtained an analytical formula for the twisting stiffness of the race shaped wires [8].

On the contrary, literature on structural behaviour of conventional slewing bearings is extensive, with several analytical models to calculate the friction torque [9–12], stiffness and static load-carrying capacity [13–15] and the effect of the preload [16–18]. Among these works, some of them aim to reduce the huge computational cost of the Finite Element Models needed to simulate the structural behaviour of the bearings; in fact, in order to properly model the contact among rolling elements and raceways, a very refined mesh is needed, and therefore the FE model becomes difficult to manage (with potential convergence problems and large analysis cost). The works of Smolnicki and Daidie were the cornerstones of those attempts [19,20].

This methodology is in line with those works aiming to reduce computational cost by replacing components of FE models with simplified mechanisms based on analytical formulations. This manuscript presents an analytical formulation, which represents the structural behaviour of the contacting elements not taking into account the flexibility of the rings, since their geometry is commonly custom-built. The methodology proposed is not purely analytical since a previous simple FE simulation must be performed to obtain contact stiffness values. In essence, the complex FE modelling of these contact zones can be substituted by the analytical formulation developed in this work, so that a much more manageable global FE model can be used to predict the structural response of the bearing.

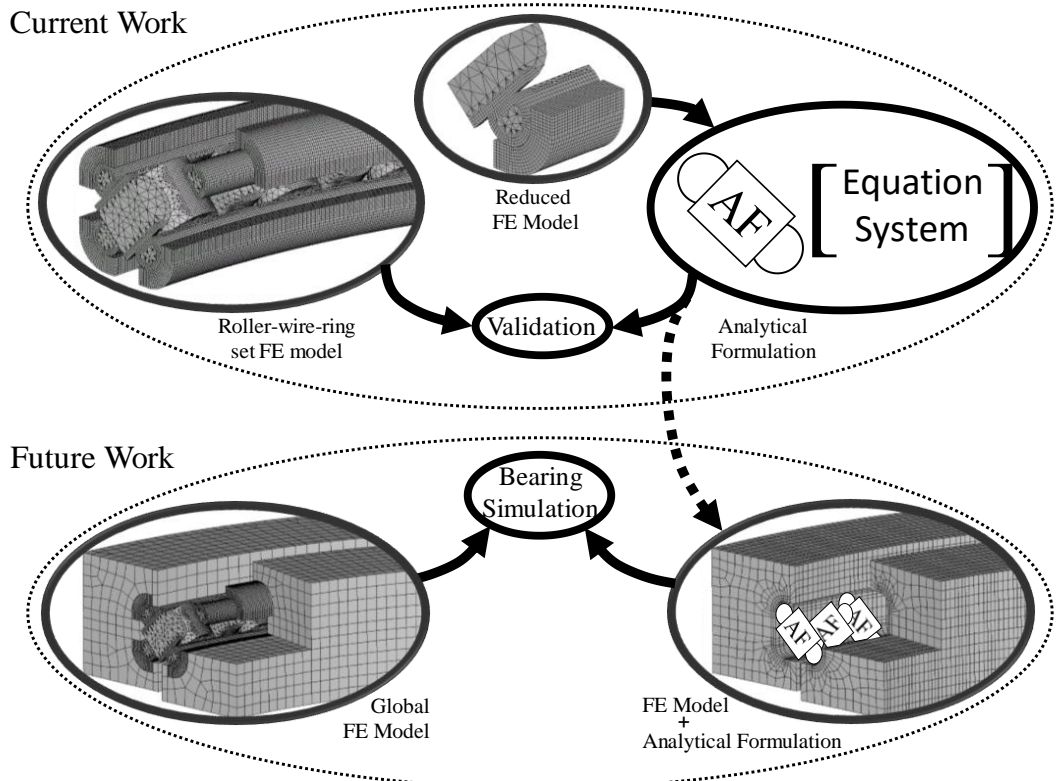

**Fig. 2.** General overview of the procedure.

## 2. Analytical submodel for the roller-wire-ring set

This section starts with the analytical formulation that represents the structural behaviour of one sector in the bearing, which corresponds to one roller-wire-ring subset, where only the local zone of the ring near the wire is considered. Then, the section continues using the previously obtained formulation to simulate the behaviour of the whole bearing, i.e. considering all the roller-wire-ring sets.

### 2.1. Analytical formulation for one sector

The aim of this formulation is to represent the structural behaviour of the contacts represented in Fig. 3, considering neither the size of the rings nor their boundary conditions, since the geometry of the rings is usually custom-built, as mentioned in the Introduction section. For the wire-ring contact, a depth of $\lambda/2$ was considered; this depth was proven to be enough to contain the local contact effects. For the development of the analytical formulation, each sector is assumed to work independently and the deformation of one sector does not affect the adjacent ones.

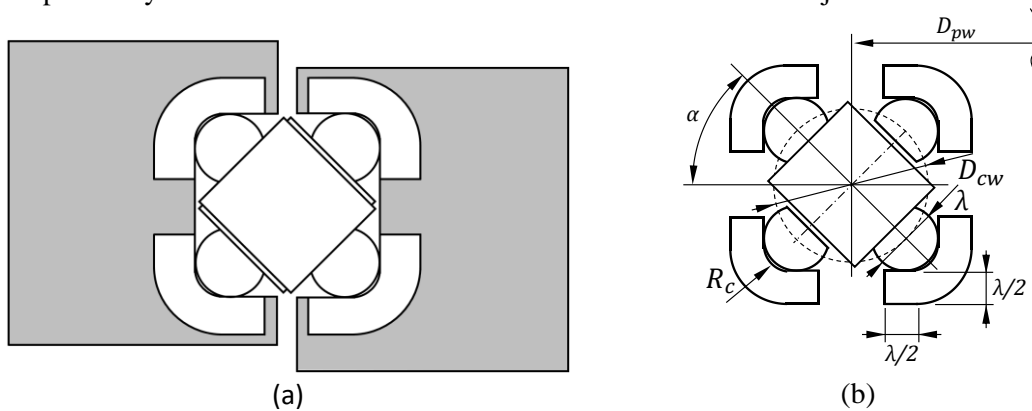

(a)          (b)

**Fig. 3.** (a) Elements considered in the analytical formulation (in grey the ones that are not represented).
(b) Geometrical parameters.

A geometrical interference model was developed to represent the structural behaviour of one sector, which is identical for roller A and roller B sectors (see Fig. 1) except for some force signs. Fig. 4(a) shows the deformed shape of a generic A sector $i$ under axial $\Delta_A^i$ and radial $\Delta_R^i$ displacements. Half displacement is applied to each ring since a symmetry point exists at the centre of the roller. The sector displacements generate contact interferences and the consequent geometrical deviations: $\Delta_1$ and $\Delta_2$ are the wire-ring interferences in the horizontal and vertical contacts, respectively, $\Delta_3$ is the roller-wire interference, and $\alpha_0$ and $\alpha$ are respectively the initial (unloaded) and final (loaded) contact angles. Regarding the reaction forces in Fig. 4(a), the model assumes that the roller-wire contact state remains "stick", and the wire-ring contact state is "slip", which allows to simulate the wire twisting phenomenon described in previous work by the authors [6]. Finally, Fig. 4(b) shows the deformation compatibility between the imposed sector displacements and the resulting contact angle and interferences.

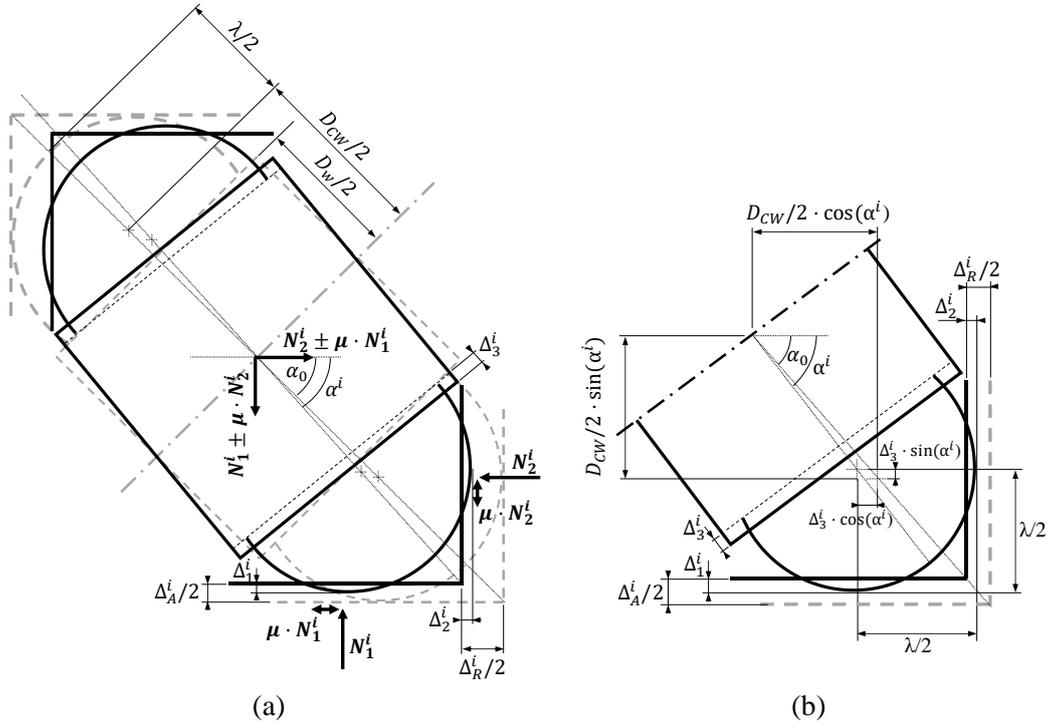

(a)        (b)
**Fig. 4.** Analytical model for roller A: (a) Deformations and forces (b) Geometrical relationships.

The direction of the friction forces in Fig. 4(a) is not trivial: it depends on the magnitude of the sector displacements $\Delta_A^i$ and $\Delta_R^i$. Under the axial displacement shown in Fig. 5 (non-deformed geometry) the wire rotates counter clockwise, but it does clockwise under the radial displacement. Obviously, the wire-ring friction reactions oppose to this rotation. In a general case with both axial and radial displacement, the dominant component must be identified and then the direction of the friction forces must be accordingly defined; thus, if $\Delta_A^i > \Delta_R^i \cdot tan(\alpha_0)$, the axial component prevails, and vice versa. If $\Delta_A^i = \Delta_R^i \cdot tan(\alpha_0)$, the friction forces must be set to zero because the wire does not twist.

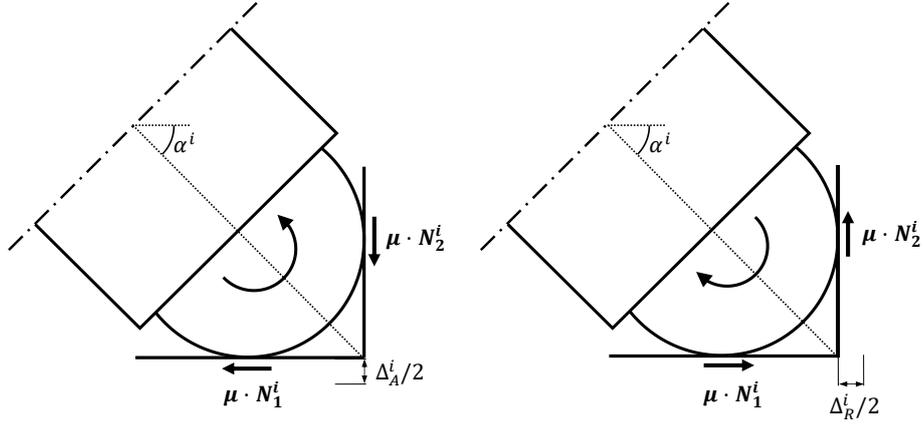

**Fig. 5.** Direction of the friction forces.

The contact force and moment equilibriums, as well as the deformation compatibility equations can now be formulated from Fig. 4:

$$N_1^i = k_1 \cdot \Delta_1^i \tag{1}$$

$$N_2^i = k_2 \cdot \Delta_2^i \tag{2}$$

$$\left(N_2^i \pm \mu \cdot N_1^i\right) \cdot cos(\alpha^i) + \left(N_1^i \pm \mu \cdot N_2^i\right) \cdot sin(\alpha^i) = k_3 \cdot \Delta_3^i \tag{3}$$

$$D_{CW}/2 \cdot cos(\alpha_0) = \Delta_R^i/2 - \Delta_2^i + \left(D_{CW}/2 - \Delta_3^i\right) \cdot cos(\alpha^i) \tag{4}$$

$$D_{CW}/2 \cdot sin(\alpha_0) = \Delta_A^i/2 - \Delta_1^i + \left(D_{CW}/2 - \Delta_3^i\right) \cdot sin(\alpha^i) \tag{5}$$

$$N_2^i \cdot \left(\left(D_{CW}/2 - \Delta_3^i\right) \cdot sin(\alpha^i)\right) \pm \mu \cdot N_1^i \cdot \left(\left((\lambda/2 - \Delta_1^i) + \left(D_{CW}/2 - \Delta_3^i\right) \cdot sin(\alpha^i)\right)\right) \pm$$
$$\pm \mu \cdot N_2^i \cdot \left((\lambda/2 - \Delta_2^i) + \left(D_{CW}/2 - \Delta_3^i\right) \cdot cos(\alpha^i)\right) - N_1^i \cdot \left(\left(D_{CW}/2 - \Delta_3^i\right) \cdot cos(\alpha^i)\right) = 0 \tag{6}$$

Where $\mu$ is the friction coefficient and $k_1, k_2, k_3$ are the stiffness constants of the roller-wire and wire-ring contacts, as shown in Fig. 4 (their values will be obtained from FE simulations as it will be explained later). $N_1, N_2$ are the wire-ring contact reactions, and the corresponding $\pm \mu \cdot N_1$ and $\pm \mu \cdot N_2$ friction forces will be positive if $\Delta_A^i > \Delta_R^i \cdot tg(\alpha_0)$ and negative otherwise.

Solving Eqs. (1-6), the results for one sector of the bearing are obtained. From those, roller-wire contact angle and normal force can be calculated:

$$F_T^i = \sqrt{\left(N_1^i \pm \mu \cdot N_2^i\right)^2 + \left(N_2^i \pm \mu \cdot N_1^i\right)^2} \tag{7}$$

$$\alpha_T^i = atan\left(\frac{\left(N_1^i \pm \mu \cdot N_2^i\right)^2}{\left(N_2^i \pm \mu \cdot N_1^i\right)^2}\right) \tag{8}$$

$$F_N^i = F_T^i \cdot cos(\alpha_T^i - \alpha^i) \tag{9}$$

### 2.2. Simulation algorithm for all the sectors

Once the analytical formulation for one sector has been explained, it must be validated. With this purpose, a simulation algorithm for all sectors was developed and programmed in Matlab®, whose results are to be compared with the results of a FE model of all the sectors explained later in this section.

The first step of the algorithm consists on distributing the amount of external displacement amongst the different roller sectors so as to solve each sector independently. In this sense, the following equations distribute the applied displacements ($\Delta_A$, $\Delta_R$, $\varphi_M$ respectively in Fig. 6) depending on the angular position of the roller sector $\theta^i$ and the mean diameter of the bearing $D_{pw}$:

$$\Delta_A^i = \Delta_A + (\varphi_M \cdot D_{pw}/2) \cdot \sin(\theta_M - \theta^i) \tag{10}$$

$$\Delta_R^i = \Delta_R \cdot \cos(\theta_R - \theta^i) \tag{11}$$

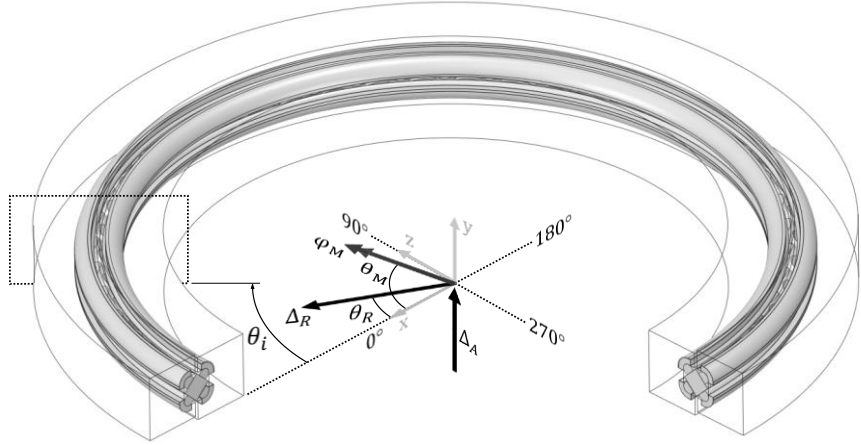

**Fig. 6.** Axial, radial and angular ring displacements

Due to the crossed layout of the rollers (see Fig. 1), each roller only contacts two wires, and under a given ($\Delta_A^i$, $\Delta_R^i$) combination the roller-wire contact may or may not be lost. Table 1 presents a summary of the different roller-wire contact possibilities under pure ring displacements $\Delta_A$, $\Delta_R$ and $\varphi_M$; in that table, rollers are classified as type A or type B depending on their orientation (see Fig. 1). Only one roller-wire contact is studied, assuming that the other contact will have the same behaviour; this is true if the inner wire and ring have approximately the same curvature as the outer ones, which is the case of wire-bearings due to their large mean diameter. For any ($\Delta_A$, $\Delta_R$, $\varphi_M$) combination, the total roller-wire contact interference can be calculated as the summation of the individual interferences in Table 1. Obviously, if the interference is positive, there exists contact between roller and wire, while a negative interference represents no contact (separation).

**Table 1**
Roller-wire contact interferences for pure ring displacements

| Axial ring displacement $\Delta_A$ | | | |
|---|---|---|---|
| **Compression direction** | | **Tension direction** | |
| **Type A roller** | **Type B roller** | **Type A roller** | **Type B roller** |
| | | | |
| $e_A^i = \Delta_A/2 \cdot \sin(\alpha_0)$ | $e_A^i = -\Delta_A/2 \cdot \sin(\alpha_0)$ | $e_A^i = -\Delta_A/2 \cdot \sin(\alpha_0)$ | $e_A^i = \Delta_A/2 \cdot \sin(\alpha_0)$ |
| Contact | Separation | Separation | Contact |
| **Radial ring displacement $\Delta_R$** | | | |
| $(\theta_R - 90°) < \theta^i < (\theta_R + 90°)$ | | $(\theta_R + 90°) < \theta^i < (\theta_R - 90°)$ | |
| **Type A roller** | **Type B roller** | **Type A roller** | **Type B roller** |
| 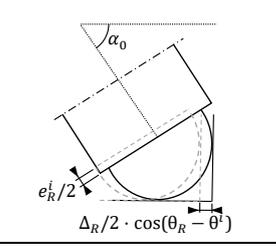 | 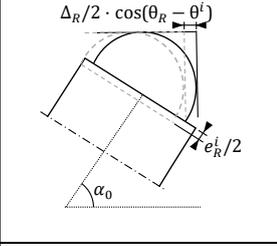 | 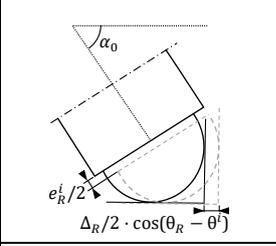 | 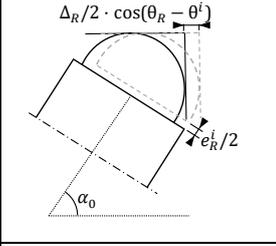 |
| $e_R^i = \Delta_R/2 \cdot \cos(\theta_R - \theta^i) \cdot \cos(\alpha_0)$ | $e_R^i = \Delta_R/2 \cdot \cos(\theta_R - \theta^i) \cdot \cos(\alpha_0)$ | $e_R^i = -\Delta_R/2 \cdot \cos(\theta_R - \theta^i) \cdot \cos(\alpha_0)$ | $e_R^i = -\Delta_R/2 \cdot \cos(\theta_R - \theta^i) \cdot \cos(\alpha_0)$ |
| Contact | Contact | Separation | Separation |
| **Angular ring displacement $\varphi_M$** | | | |
| $(\theta_M - 180°) < \theta^i < (\theta_M)$ | | $(\theta_M) < \theta^i < (\theta_M + 180°)$ | |
| **Type A roller** | **Type B roller** | **Type A roller** | **Type B roller** |
| 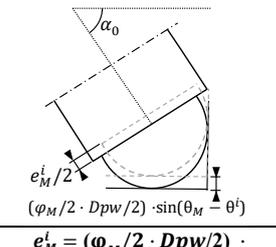 | 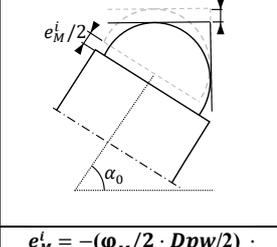 | 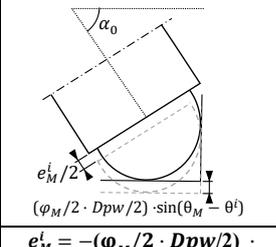 | 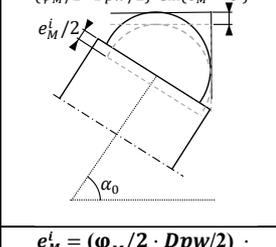 |
| $e_M^i = (\varphi_M/2 \cdot Dpw/2) \cdot \sin(\theta_M - \theta^i) \cdot \sin(\alpha_0)$ | $e_M^i = -(\varphi_M/2 \cdot Dpw/2) \cdot \sin(\theta_M - \theta^i) \cdot \sin(\alpha_0)$ | $e_M^i = -(\varphi_M/2 \cdot Dpw/2) \cdot \sin(\theta_M - \theta^i) \cdot \sin(\alpha_0)$ | $e_M^i = (\varphi_M/2 \cdot Dpw/2) \cdot \sin(\theta_M - \theta^i) \cdot \sin(\alpha_0)$ |
| Contact | Separation | Separation | Contact |

Once the contact status for each roller is determined, the next step consists in solving each sector with the analytical formulation and the corresponding $(\Delta_A^i, \Delta_R^i)$ displacements. In order to obtain the total forces and moments in the axis from the Fig. 6, the following formulas must be applied:

$$F_x^\square = \sum_{rollers\ A} (N_{2A}^i \pm \mu \cdot N_{1A}^i) \cdot \cos(\theta^i) + \sum_{rollers\ B} (N_{2B}^i \pm \mu \cdot N_{1B}^i) \cdot \cos(\theta^i) \quad (9)$$

$$F_y^\square = \sum_{rollers\ A} (N_{1A}^i \pm \mu \cdot N_{2A}^i) - \sum_{rollers\ B} (N_{1B}^i \pm \mu \cdot N_{2B}^i) \quad (10)$$

$$F_z^\square = \sum_{rollers\ A} (N_{2A}^i \pm \mu \cdot N_{1A}^i) \cdot \sin(\theta^i) + \sum_{rollers\ B} (N_{2B}^i \pm \mu \cdot N_{1B}^i) \cdot \sin(\theta^i) \quad (11)$$

$$M_z^\square = M_{zA}^\square + M_{zB}^\square$$

$$M_{zA}^\square = \sum_{rollers\ A} \left( N_{1A}^i \cdot \left(\frac{D_{pw}}{2} - \frac{D_{cw}}{2} \cdot \cos(\alpha^i)\right) \cdot \cos(\theta^i) \pm \mu \cdot N_{1A}^i \cdot \cos(\theta^i) \cdot \left(\frac{D_{cw}}{2} \cdot \sin(\alpha^i) + \frac{\lambda}{2}\right) + \right.$$
$$\left. + N_{2A}^i \cdot \cos(\theta^i) \cdot \left(\frac{D_{cw}}{2} \cdot \sin(\alpha^i)\right) \pm \mu \cdot N_{2A}^i \cdot \left(\frac{D_{pw}}{2} - \frac{D_{cw}}{2} \cdot \cos(\alpha^i) - \frac{\lambda}{2}\right) \cdot \cos(\theta^i) \right)$$

$$M_{zB}^\square = \sum_{rollers\ B} \left( -N_{1B}^i \cdot \left(\frac{D_{pw}}{2} - \frac{D_{cw}}{2} \cdot \cos(\alpha^i)\right) \cdot \cos(\theta^i) \pm \mu \cdot N_{1B}^i \cdot \cos(\theta^i) \cdot \left(\frac{D_{cw}}{2} \cdot \sin(\alpha^i) + \frac{\lambda}{2}\right) - \right.$$
$$\left. - N_{2B}^i \cdot \cos(\theta^i) \cdot \left(\frac{D_{cw}}{2} \cdot \sin(\alpha^i)\right) \pm \mu \cdot N_{2B}^i \cdot \left(\frac{D_{pw}}{2} - \frac{D_{cw}}{2} \cdot \cos(\alpha^i) - \frac{\lambda}{2}\right) \cdot \cos(\theta^i) \right) \quad (12)$$

$$M_z^\square = M_{zA}^\square + M_{zB}^\square$$

$$M_{xA}^\square = \sum_{rollers\ A} \left( -N_{1A}^i \cdot \left(\frac{D_{pw}}{2} - \frac{D_{cw}}{2} \cdot \cos(\alpha^i)\right) \cdot \sin(\theta^i) \pm \mu \cdot N_{1A}^i \cdot \sin(\theta^i) \cdot \left(\left(\frac{D_{cw}}{2} \cdot \sin(\alpha^i)\right)\right) - \right.$$
$$\left. - N_{2A}^i \cdot \sin(\theta^i) \cdot \left(\frac{D_{cw}}{2} \cdot \sin(\alpha^i)\right) \pm \mu \cdot N_{2A}^i \cdot \left(\frac{D_{pw}}{2} - \frac{D_{cw}}{2} \cdot \cos(\alpha^i) - \frac{\lambda}{2}\right) \cdot \sin(\theta^i) \right) \quad (13)$$

$$M_{xB}^\square = \sum_{rollers\ B} \left( N_{1B}^i \cdot \left(\frac{D_{pw}}{2} - \frac{D_{cw}}{2} \cdot \cos(\alpha^i)\right) \cdot \sin(\theta^i) \pm \mu \cdot N_{1B}^i \cdot \sin(\theta^i) \cdot \left(\left(\frac{D_{cw}}{2} \cdot \sin(\alpha^i)\right)\right) + \right.$$

$$+N_{2B}^i \cdot sin(\theta^i) \cdot \left(\frac{D_{cw}}{2} \cdot sin(\alpha^i)\right) \pm \mu \cdot N_{2B}^i \cdot \left(\frac{D_{pw}}{2} - \frac{D_{cw}}{2} \cdot cos(\alpha^i) - \frac{\lambda}{2}\right) \cdot sin(\theta^i)\right)$$

## 3. Finite Element Models

Two different Finite Element (FE) models were built in this work. The first one is a simplified model to obtain the contact stiffness constants for the analytical model. The second FE model is a detailed model of what the analytical formulation represents, developed with validation purposes.

### 3.1. FE model to obtain the contact stiffness constants

In principle, some Hertz-type analytical approaches from literature could be used to estimate the roller-wire and wire-ring contact stiffness values, but they are based on plane strain or plane stress assumptions, which is not the case of crossed roller wire bearings. Thus, FE analyses were used to accurately obtain the values of the roller-wire and wire-ring contact stiffnesses $k_1, k_2$ and $k_3$, which must be introduced in Eqs. (1-6) of the analytical formulation. Fig. 7(a) shows the simplified model, where only one eighth of a sector was modelled taking advantage of the cyclic symmetry and symmetry planes of the system. Several partitions were carried out in order to achieve a highly efficient mesh. As usual in wire bearings, the material of roller and wire is steel, and a lighter material is commonly used for the ring. All contacts were defined as frictional ($\mu$=0.1) with a penetration tolerance of 0.1μm in Ansys®.

As mentioned before, the analytical model only accounts for contact deformations, not considering the flexibility of the rings. To simulate this condition in the FE model, the external faces of the ring partition were clamped and symmetry boundary conditions were applied to the cyclic symmetry cutting faces. Finally, the load was applied to the roller as a displacement δ in the direction of the contact angle (see Fig. 7).

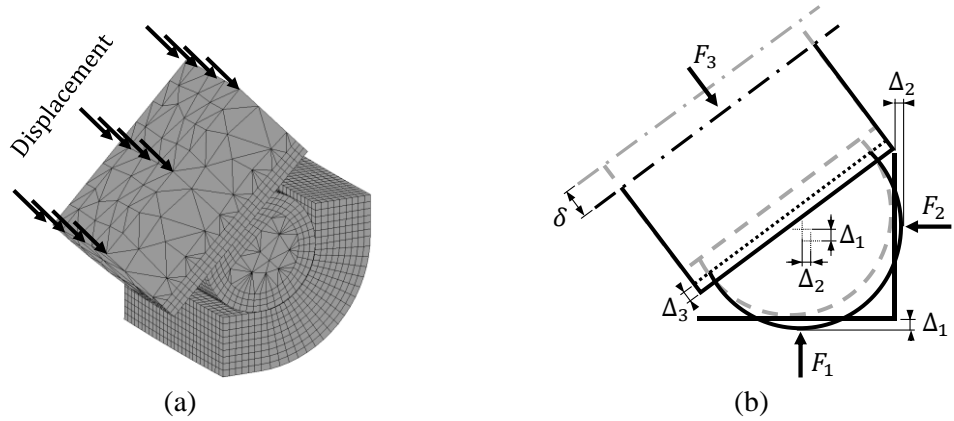

**Fig. 7.**: (a) Reduced FE model (b) Forces and displacements in the reduced FE model.

The stiffness for each contact, which happens to be constant according to the analyses performed during the development of this work, is calculated from the contact deformations and reaction forces, as indicated in Fig. 7(b). Reaction forces are directly obtained from the FE results, while the deformations can be obtained from the relative displacement of the wire centre:

$$k_1 = F_1/\Delta_1 \qquad (9)$$
$$k_2 = F_2/\Delta_2 \qquad (10)$$

$$k_3 = F_3/\Delta_3 = \frac{F_3}{\delta - \sqrt{\Delta_1^2 + \Delta_1^2}} \tag{11}$$

As a further step, a DoE (Design of Experiments) with different geometries and materials could be performed and the values for the stiffness constants $k_1$, $k_2$ and $k_3$ could then formulated in terms of the geometrical dimensions and parameters under study. In this sense, using those expressions the developed methodology would become fully analytical.

### 3.2. FE model to validate the analytical model

The aim of this FE model is to analyse pure axial, radial and moment load cases to compare its results with the ones provided by the analytical algorithm. Only half of the bearing is modelled taking advantage of the symmetry plane, as shown in Fig. 8 (a).

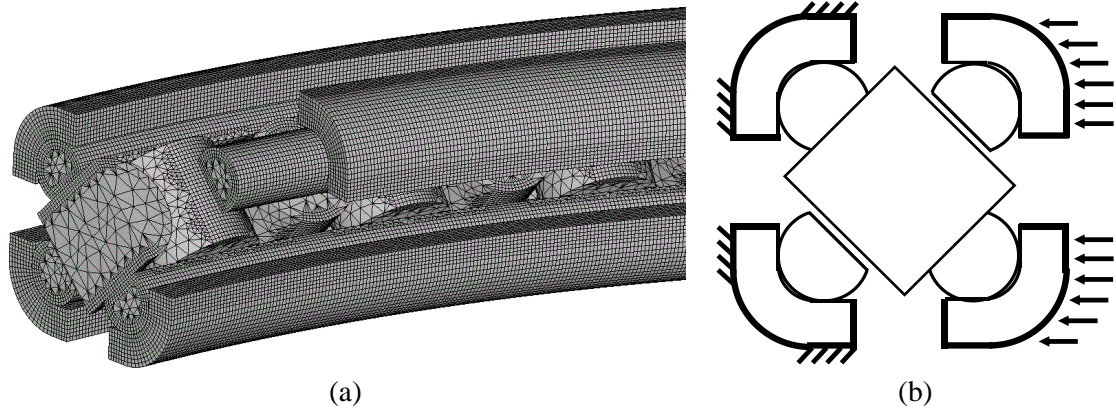

(a)                      (b)

**Fig. 8.** (a) Local FE model (11.749.170 DoF) (b) Boundary conditions under radial load

Regarding the mesh, the geometry of the wires and rollers was divided into several partitions that enable to perform a finer mesh in the contact regions. A sensitivity study of the mesh reported this strategy to be optimal, providing elements with good aspect ratio and abruptly reducing the computational cost without compromising the accuracy of the results. The contacts were defined as frictional (μ=0.1) with a penetration tolerance of 0.1μm in the roller-wire and 0.5μm in the wire-ring interfaces.

Again, the external faces of the ring partitions were rigid in order to simulate the analytical formulation. Finally, ring displacements were introduced by means of a remote node located in the centre of the bearing and rigidly connected to the internal faces of the inner ring partitions, while the outer ring partitions were fixed (see Fig. 8(b) for the radial case).

### 4. Results and Discussion

Next, the results of the analytical algorithm and FE model explained in the previous sections are presented and compared in order to validate the analytical model developed in this work. Three pure load cases were analysed: axial force, radial force and tilting moment, until the static load-carrying capacity, which corresponds to the load that causes a roller-wire contact pressure of 4000 MPa, according to ISO standard for conventional bearings [21]. To improve convergence, displacements were applied instead of forces, reporting the reactions as the corresponding load magnitudes. Although the half bearing model in Fig. 8 was used for the radial and moment load cases, only half sector was analysed (with two roller halves) in the axial case due to the cyclic symmetry of the problem.

Table 2 points out the geometrical parameters of the analysed bearing (see Fig. 2(b)). These values were obtained from commercial catalogues [22]. Regarding the material of the rings, aluminium was selected since it is the most usual material for these bearings.

**Table 2**
Geometrical parameters

| $D_w$ | $D_{pw}$ | $D_{cw}$ | $R_c$ | $\lambda$ | $\alpha$ | $N$ |
|---|---|---|---|---|---|---|
| 14 mm | 420 mm | 18 mm | 3.9 mm | 8 mm | 45° | 94 |

Regarding the contact stiffness constants necessary for the analytical formulation to work, a FE analysis with the model of the Fig. 7(a) and the proper geometrical parameters was performed. The FE force-deformation results of each contact are presented in Fig. 9 together with the linear approximations that retrieves the contact stiffness constants of Table 3.

**Table 3**
Contact stiffness constants

| $k_1$ | $k_2$ | $k_3$ |
|---|---|---|
| 372509 [N/mm] | 368393 [N/mm] | 447544 [N/mm] |

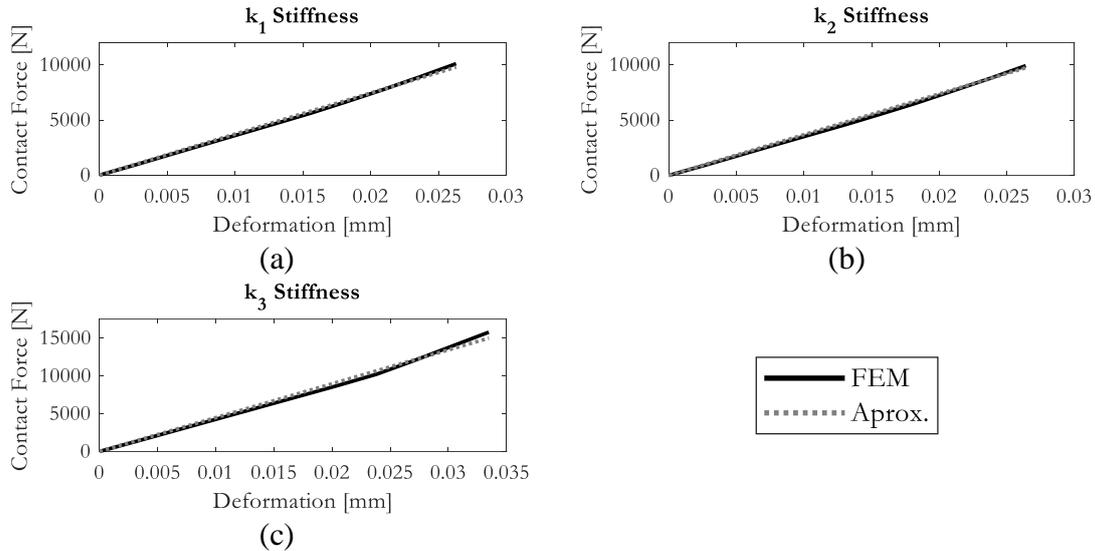

**Fig. 9.** Contact stiffness according FE results: (a) $k_1$ (b) $k_2$ (c) $k_3$

### 4.1. Stiffness of the roller-wire-ring set

The relationship between the applied load and the resulting deformation defines the stiffness of the system. Stiffnesses under pure axial, radial and tilting loads are a key parameter in the bearing design and selection for different applications. In this sense, Fig. 9 shows the analytical and FE results for the geometry defined in Table 2.

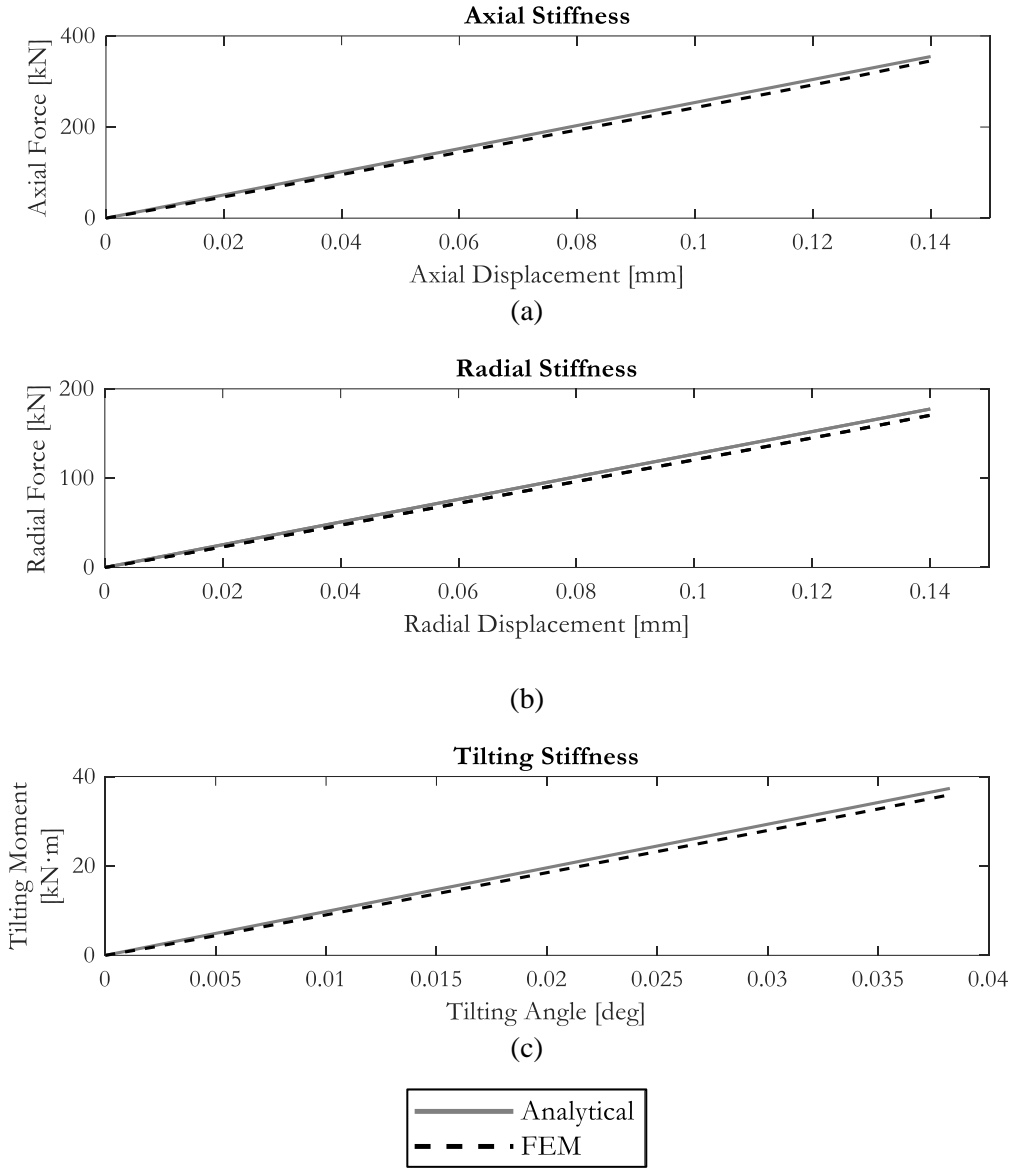

**Fig. 9.** Bearing stiffnesses: (a) Axial stiffness (b) Radial stiffness (c) Tilting stiffness

It can be observed that the correlation between the FE analyses and the algorithm is excellent, The behaviour in this particular cases is almost linear, so the approximated stiffness constant for each case was obtained and shown in Table 3, where it can also be appreciated that the error is minimum.

**Table 3**
Stiffness constants of the validation and relative errors

| Load Case | FE model | Analytical Formulation | Error % |
|---|---|---|---|
| Axial [kN/mm] | 2437 | 2538 | 4.1 |
| Radial [kN/mm] | 1208 | 1269 | 5.0 |
| Moment [kN·m/deg] | 930779 | 977518 | 5.0 |

### 4.2. Normal contact force and angle

The normal contact angle defines the final position of the wires-roller set, which may rotate with respect to the rings as explained before. Fig. 10 shows that the analytical and FE results are in excellent agreement regarding normal contact forces and angles. Note that apart from predicting the load distribution, the analytical model also represents correctly the wire rotation phenomenon illustrated in Fig. 5, since the contact angle decreases in the axial and tilting load cases and increases in the radial load case.

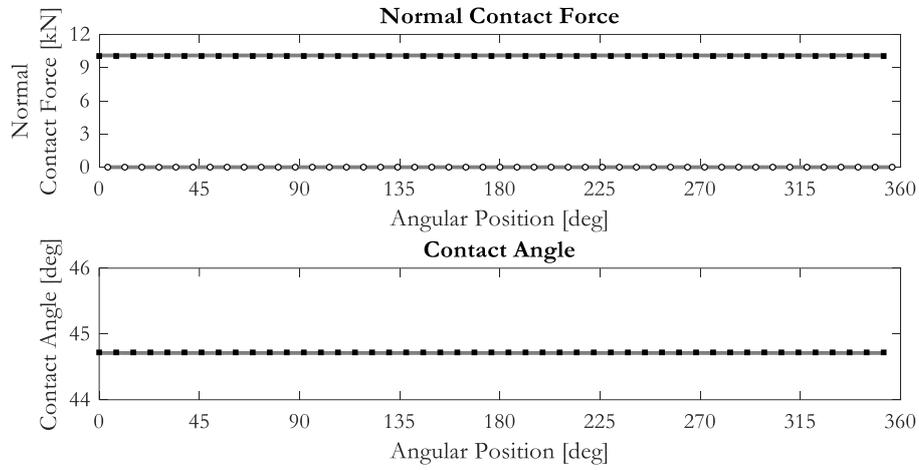

(a)

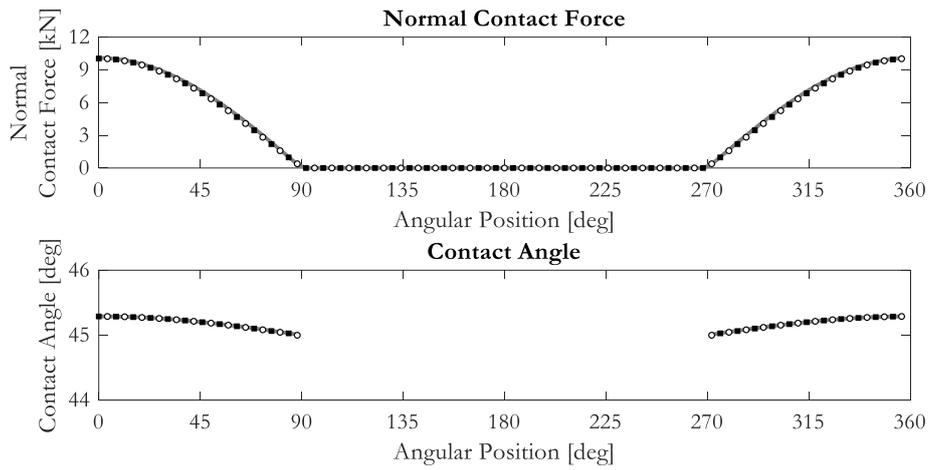

(b)

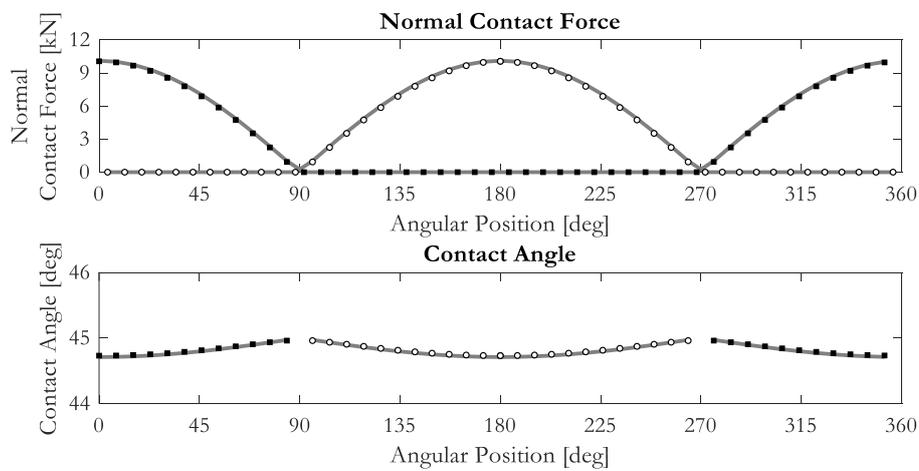

(c)

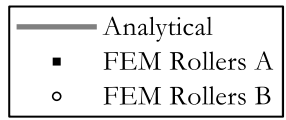

**Fig. 10.** Contact normal force and angle for the maximum load: (a) Axial load (b) Radial load (c) Tilting moment load

## 5. Conclusions and Future Work

An analytical formulation that simulates the static structural behaviour of the local geometry of the roller-wire-ring set has been developed. Based on a contact interference model, it reproduces the interaction between roller, wires and rings, estimating the contact forces and deformations. To validate the formulation, it was implemented in a simulation algorithm where the stiffness, contact forces and contact angles were calculated and compared with FE simulations. Overall, the results reveal that the simulation algorithm accurately estimates the load distribution and stiffness under axial, radial and tilting loads. The errors in stiffness results are slightly larger but negligible in any case.

This simplified model can be implemented in global Finite Element models with their particular ring geometries and boundary conditions (which can differ greatly depending on the application) so that the complexity, cost and potential convergence problems of the FE analysis are greatly reduced. In this sense, further research will implement this methodology into a global FE model with its corresponding ring geometry and boundary conditions.

### Acknowledgements

The authors want to acknowledge the financial support of the Spanish Ministry of Economy and Competitiveness through grant number DPI2017-85487-R (AEI/FEDER,UE); the Basque Government through project number IT947-16 and research program HAZITEK 2019, acronym WIRE; and the University of the Basque Country through doctoral grant number PIF18/071.

Special thanks also to Basque manufacturer Iraundi Bearings S.A. for the help in many background items of the research.